\documentclass[aps,pre,a4paper,12pt]{revtex4-1}%

\pdfoutput=1
\usepackage{amssymb}
\usepackage{amsmath}
\usepackage{mathrsfs}
\usepackage{bm}
\usepackage[dvipsnames]{xcolor}
\usepackage{graphicx}
\usepackage{color}
\usepackage{mathtools}

\begin{document}

\title{Polymerization induces non-Gaussian diffusion}

\author{Fulvio Baldovin}
%\affiliation{Dipartimento di Fisica and Sezione INFN, Università degli Studi di Padova, I-35131 Padova, Italy}
\author{Enzo Orlandini}
%\affiliation{Dipartimento di Fisica and Sezione INFN, Università degli Studi di Padova, I-35131 Padova, Italy}
\author{Flavio Seno}
\affiliation{Dipartimento di Fisica and Sezione INFN, Università degli Studi di Padova, I-35131 Padova, Italy}
%\date{\today}

\begin{abstract}
Recent theoretical modeling offers a unified picture for the
description of stochastic processes characterized by a crossover from
anomalous to normal behavior. This is particularly welcome,
as a growing number of experiments suggest the crossover to be a
common feature shared by many systems: in some cases the anomalous
part of the dynamics amounts to a Brownian yet non-Gaussian diffusion;
more generally, both the diffusion exponent and the distribution may
deviate from normal behavior in the initial part of the process.
Since proposed theories work at a mesoscopic scale invoking the
subordination of diffusivities, it is of primary importance to bridge
these representations with a more fundamental, ``microscopic''
description. We argue that the dynamical behavior of macromolecules
during simple polymerization processes provide suitable setups in which analytic,
numerical, and particle-tracking experiments can be
contrasted at such a scope. Specifically, we demonstrate that Brownian
yet non-Gaussian diffusion of the center of mass of a polymer is a
direct consequence of the polymerization process. Through the
kurtosis, we characterize the early-stage non-Gaussian behavior within
a phase diagram, and we also put forward an estimation for the crossover
time to ordinary Brownian motion. 
\end{abstract}

\maketitle
\section{Introduction}
Diffusion in crowded and complex systems such as biological cells is
usually very heterogeneous, and  anomalous behavior -- where the mean
square displacement of tracers varies non linearly with time -- is
envisaged~\cite{metzler2000,sanabria2007,hofling2013}.
Over the last few
years a new class of diffusive processes has been reported, where the
mean square displacement
is found to grow linearly in time like in standard, Brownian
diffusion, but with a corresponding probability density function (PDF)
which is strongly non-Gaussian~\cite{weeks2000,hapca2008,wang2009,toyota2011,wang2012,guan2014,ghosh2014,wang2014,stylianidou2014,samanta2016,dutta2016,metzler2017,cherstvy2018}.
This behavior, termed Brownian yet non-Gaussian 
diffusion~\cite{wang2009,wang2012}, occurs quite robustly in a 
wide range of systems, including beads diffusing on lipid tubes~\cite{wang2009} or in
networks~\cite{wang2009,toyota2011},  the motion of tracers in colloidal, polymeric or active 
suspensions~\cite{weeks2000,kegel2000,leptos2009,xue2016}
and in biological cells\cite{stylianidou2014,parry2014,munder2016}, as well as the motion of individuals in
heterogeneous populations such as nematodes~\cite{hapca2008}. Similar effects on the PDF
are also observed in the anomalous diffusion~\cite{lampo2017} of labeled
messenger RNA molecules in living $E. coli$ and $S. cervisiae$ cells.
In the majority of cases, at larger time the form of the PDF crosses over to the 
normal, Gaussian one. Therefore, such change cannot be simply
due to the heterogeneity of the tracers, unless some of their properties
vary with time. More plausibly, the anomalous-to-Gaussian transition
might be induced by temporal fluctuations of the diffusion
coefficient, due to rearrangements of properties of tracers or of the
surrounding medium. To mimic such behaviors, models in which the
diffusion varies with time by obeying a stochastic equation has been
introduced and solved both analytically than numerically. These models
are referred in the literature as the ``diffusing diffusivity models''~\cite{chubynsky2014,chechkin2017,jain2017,jain2017b,tyagi2017,matse2017,jain2018,sposini2018,sposini2018b,grebenkov2019}, and it
has been shown that for short times they are intimately related to the
idea of superstatistics~\cite{beck2003}.
%, an approached promoted by Beck and Cohen\cite{beck2003}. 
In the latter approach, an ensemble of particles is assumed to be characterized by
different diffusion coefficients and it is then described as a mixture of Gaussian
PDFs, weighted by the distribution of the diffusivities. 
%Infact, for an ensemble of particles, we could
%immagine that the non Gaussian statistic  emerges because
%different particles are located in different enviroments,
%each characterized by a diffusion coefficients $D$. 
%If during the observation time each particle diffuses in its own enviroment,  
%the statistics of the 
%ensemble of particles can be described as a mixture of individual Gaussians, weighted by some distribution $p(D)$
%of local diffusivities $D$.
As a result, the ensemble dynamics is still
Brownian, yet the PDF of particle displacements corresponds to a
Gaussian mixture and it is thus
%to a sum of single Gaussians with specific value of $D$, weighted by $p(D)$, i.e. a  
%distribution that is
not  Gaussian  anymore.

Although diffusing diffusivity models
qualitatively reproduce the experimental observations, they
work at a mesoscopic scale and without a visible connection to the
underlying molecular processes. It is therefore becoming increasingly
relevant to find a strategy that bridges the gap between the paradigm of
diffusing diffusivity and the microscopic realm, in order to fully
understand this form of anomalous diffusion.  In this paper we
show how the diffusion of polymers during a polymerization process
offers one possible mechanism to realize this connection.  It is well
known from polymer theory~\cite{doi1988} that the motion of the
center of mass of a linear chain is Brownian, but with a diffusivity
constant which is inversely proportional to $N^{\alpha}$, where $N$ is the
number of monomers and $\alpha$ an exponent ranging from $1/2$
(Rouse model) to $2$ (reptation model). During 
an equilibrated polymerization processes the number $N$ fluctuates in time and its 
statistics can be obtained through the exact solution of its stationary
master equation. By using a continuous approximation for this
temporally homogeneous birth-death Markov
process~\cite{gillespie1992}, it emerges
that in the limit of large systems such process converges to an
Ornstein-Uhlenbeck, as it is assumed in most of the diffusing
diffusivity models~\cite{chechkin2017}.
%Notably,
%the resulting equilibrium diffusion for diffusivities and the PDF of the displacements, that can
%be deduced by using the superstatistical formalism, is
%different by those imposed in diffusing diffusivity models; in
%particular a power law behavior of the PDF for long distances is observed.
The time scale of the Ornstein-Uhlenbeck process is linearly proportional to
the volume of the system and this guarantees that the non-Gaussian
behavior can be accessible experimentally by tuning such parameter.

%\section{Tagged monomer dynamics}

\section{Polymerization process}
Polymers are made of relatively simple subunits (monomers)
assembled with one another through different
mechanisms and geometries. The result is a macromolecule which may contain
from a few tens (in the case oligomers), to several
thousand monomer units~\cite{flory1942}, or even millions as in the case of DNA and RNA molecules.
From a biological point of view, the polymerization process occurs regularly either within or outside the
cell~\cite{paul2012} . In particular, cells might trigger
polymerization by several mechanisms such as the \textit{de novo}
nucleation of new filaments, the uncapping of existing barbed ends
(actin) and rescuing a depolymerizing filament (commonly observed for
microtubules).

In order to guarantee the existence of equilibrium conditions,
here we  consider a polymerization process occurring in a closed volume 
with a fixed total number of  monomers $N_{\mathrm{t}}$.
For sake of simplicity, in what follows we suppose
that one filament only can nucleate and that subunits may  bind
reversibly onto both ends of the chain.  At each end, the addition
and deletion of monomers can be represented as~\cite{boal2002}
\begin{equation}
A_N+A_1\xrightleftharpoons[k_-]{k_+} A_{N+1}\,,
\end{equation}
where $A_N$ is the filament with $N$ subunits, and $k_+$, $k_-$ are
the rate constants for association and dissociation, respectively.
Hence,
\begin{equation}
N_{\mathrm{t}}=N(t)+M(t)\,,
\end{equation}
where $M(t)=c(t)\,V$ is the number of monomeric subunits, $c$ its
concentration and $V$ the system volume. The probability of a filament
with $n$ monomers at time $t$ given $n_0$ units at time $t_0$,
$P_N(n,t|n_0,t_0)$ satisfies the (forward) master equation of a
temporally homogeneous birth-death Markov
process~\cite{gillespie1992}:
\begin{equation}
  \begin{array}{lll}
     \partial_tP_N(n,t|n_0,t_0)&=&
    \left[W_-(n+1)\,P_N(n+1,t|n_0,t_0)
    -W_+(n)\,P_N(n,t|n_0,t_0)\right]\\
    &&+\left[W_+(n-1)\,P_N(n-1,t|n_0,t_0)
    -W_-(n)\,P_N(n,t|n_0,t_0)\right]
  \end{array}\,,
  \label{eq_master}
\end{equation}
with stepping functions
\begin{equation}
  \begin{array}{l}
    W_+(n)=2k_+\,c(n)\quad(1\leq n\leq N_{\mathrm{t}})\,,
   \\
    W_-(1)=0\,,
    \quad W_-(2)=k_-\,,
    \quad W_-(n)=2k_-\quad(3\leq n\leq N_{\mathrm{t}})\,,
  \end{array}
  \label{eq_stepping}
\end{equation}
and $c(n)=(N_{\mathrm{t}}-n)/V$.
Through these choices, we are assuming with certainty the existence in
solution of a filament with at least one monomer.  The factor $2$ in
$W_+$ models a linear polymer which grows at both
ends without developing branching;
$W_-$ is instead concerned with the 
possible bonds which may break down. 
%and $S(N)$ is a surface factor expressing
%the number of polymer's subunits available to association and
%dissociation. For a linear polymer, $S(N)=2$ (independently on $N$),
%in view of the 
%filament's two ends; for a globular compact polymer,
%$S(N)\sim N^{2/3}$, since in this case the polymer radius grows with
%$N$ as $R\sim N^{1/3}$ . 
Equilibrium is reached under detailed balance $W_-(n)=W_+(n)$
($3\leq n\leq N_{\mathrm{t}}$), corresponding to  a
polymer composed by
\begin{equation}
N_{\mathrm{eq}}=N_{\mathrm{t}}-\frac{k_-}{k_+}\,V\equiv\lambda\,N_{\mathrm{t}}
\end{equation}
monomers, and to a number 
\begin{equation}
M_{\mathrm{eq}}=\frac{k_-}{k_+}\,V\equiv(1-\lambda)\,N_{\mathrm{t}}
\end{equation}
of single monomers in solution.
We remark that the rate
constants $k_+$, $k_-$ are specific to the polymerization chemical reactions.  
Given a certain kind of polymer,
the average polymer size and the average 
number of single monomers in solution  are thus
controlled by the total number of subunits
$N_{\mathrm{t}}$ and  by the volume of the system $V$, which are
quantities easily controlled in experiments.
In the following analysis, we find it convenient to replace the volume
with the fraction $0<\lambda<1$ of $N_{\mathrm{t}}$ that
compose the polymer at equilibrium; clearly,
$V=(1-\lambda)\,N_{\mathrm{t}}\,k_+/k_-$. 

As we prove in the Appendix, for any given $N_{\mathrm{t}}$ and
independently from $n_0$, the 
stationary solution $P_N(n)\equiv\lim_{t\to\infty}P_N(n,t|n_0,t_0)$ reads
\begin{equation}
  \begin{array}{lll}
    P_N(1)
    &=&
    \displaystyle
    \dfrac{1}{\mathcal{N}(N_{\mathrm{t}},\lambda)}
    \;\dfrac{(1-\lambda)\,N_{\mathrm{t}}}{2\,(N_{\mathrm{t}}-1)}
    \\
    &&\\
    P_N(2)
    &=&
    \displaystyle
    \dfrac{1}{\mathcal{N}(N_{\mathrm{t}},\lambda)}
    \\
    &&\\
    P_N(n)
    &=&
    \displaystyle
    \dfrac{2}{\mathcal{N}(N_{\mathrm{t}},\lambda)}
    \;\frac{(N_{\mathrm{t}}-2)!}{\left[(1-\lambda)\,N_{\mathrm{t}}\right]^{N_{\mathrm{t}}-2}}
    \;\frac{\left[(1-\lambda)\,N_{\mathrm{t}}\right]^{N_{\mathrm{t}}-n}}{(N_{\mathrm{t}}-n)!}
    \quad(3\leq n\leq N_{\mathrm{t}})
  \end{array}\,,  
  \label{eq_p_N_master}
\end{equation}
with a normalization factor
\begin{eqnarray}
  \mathcal{N}(N_{\mathrm{t}},\lambda)
  &=&
  \frac{N_{\mathrm{t}}\,[(11-4\lambda)\,\lambda-9]+2}{2(N_{\mathrm{t}}-1)}
  \nonumber\\
  &&
  +\frac{2\,(N_{\mathrm{t}}-2)!}{\left[(1-\lambda)\,N_{\mathrm{t}}\right]^{N_{\mathrm{t}}-2}}
  \;\frac{\Gamma(N_{\mathrm{t}}+1,(1-\lambda)\,N_{\mathrm{t}})}{\Gamma(N_{\mathrm{t}}+1)}
  \;\mathrm{e}^{(1-\lambda)\,N_{\mathrm{t}}}\,,
  \label{eq_normalization}
\end{eqnarray}
$\Gamma(\cdot,\cdot)$ being the upper incomplete gamma
function~\cite{abramowitz1965},
\begin{equation}
  \Gamma(N_{\mathrm{t}}+1,(1-\lambda)\,N_{\mathrm{t}})
  \equiv\int_{(1-\lambda)\,N_{\mathrm{t}}}^{\infty}\mathrm{d}t
  \,t^{N_{\mathrm{t}}}\,\mathrm{e}^{-t}\,,
\end{equation} 
and $\Gamma(\cdot)$ the Euler gamma function. We may observe that
with $(1-\lambda)\,N_{\mathrm{t}}\to0$ the two Gamma functions in the
normalization factor become equal and simplify to $1$;
in this limit, probabilities for small $n$ are suppressed.
Indeed, in Section~\ref{section_crossover} we show that
$P_N(n)$ becomes close to a Gaussian
for large $\lambda$ and $N_{\mathrm{t}}$.
In view of the
inverse power-law relation with the diffusion coefficient of the center of mass,
it is however the behavior for  small $n$ which affects the probability of 
large diffusivities, triggering in turn strong deviations from ordinary
diffusion which are described in the following Section.

\section{Brownian yet non-Gaussian diffusion of the center of mass}
From polymer physics we know that the center of mass
$\boldsymbol{R}_G$ of a macromolecules  with $N$ subunits  
diffuses with a coefficient $D(N)=D_0/N^\alpha$, $D_0$ being a
diffusion coefficient specific of the  considered subunit. This means
\begin{equation}
\mathrm{d}\boldsymbol{R}_G(t)=\sqrt{6\,D(N(t))}\,\mathrm{d}\boldsymbol{B}(t)\,,
\end{equation}
with $\boldsymbol{B}(t)$  a (three-dimensional) Wiener process (Brownian motion). 
Reference values for the exponent $\alpha$ are:
\begin{itemize}
\item $\alpha=1/2$ in the Rouse
  model~\cite{doi1988,weber2010}, where the polymer is composed of $N$
  equivalent beads with neither
  excluded-volume nor hydrodynamic interaction;
\item $\alpha=1$ for the Zimm
  model~\cite{ermak1978,doi1988}, where hydrodynamic is taken into
  account;
\item $\alpha=2$ for the reptation model which 
  describes tagged polymer motion in entangled polymer
  solutions~\cite{doi1978,doi1988}.
\end{itemize}
In view of the previous analysis, we
understand that polymerization confers a random character
to  $\boldsymbol{R}_G$, providing a clear microscopic origin to the
``diffusing diffusivity'' process we are going to detail next.

From Eq.~\eqref{eq_p_N_master} we 
readily obtain the stationary distribution for the diffusion
coefficient of the polymer's center of mass, 
\begin{equation}
  P_{D}(D_n)
  =\sum_{n'=1}^{N_{\mathrm{t}}}P_{N}(n')\;\delta_{D_n,\frac{D_0}{n'^\alpha}}
  =P_{N}\left(\dfrac{D_0^\alpha}{D_n^\alpha}\right)
  \quad(1\leq n\leq N_{\mathrm{t}},\;D_n=D_0/n^\alpha)\,,
  \label{eq_p_D_1}
\end{equation}
and its first moment
\begin{equation}
  D_{\mathrm{av}}\equiv\mathbb{E}[D_n]=\sum_{n=1}^{N_{\mathrm{t}}}
  P_{D}(D_n)\;D_n\,.
\end{equation}
Imagine now to perform a particle-tracking experiment at constant $N_{\mathrm{t}}$ and $V$ and to 
monitor the  position of $\boldsymbol{R}_G$ in stationary conditions.
At a given initial instant the polymer possesses a size $n$, and thus
a diffusion coefficient $D_n=D_0/n^\alpha$ with
probability given by Eq.~\eqref{eq_p_D_1}.
For time smaller than the characteristic decay $\tau$ of the autocorrelation of the process
$N(t)$, the experimental PDF amounts then to a Gaussian
mixture (also called ``superstatistics'')~\cite{wang2009,chubynsky2014,beck2003}
weighted by Eq.~\eqref{eq_p_D_1}. In addition, its second moment grows
linearly with time as in the ordinary Brownian motion. Such a
phenomenon of ``Brownian yet non Gaussian
diffusion''~\cite{wang2009,wang2012} has been recently modeled at
a mesoscopic scale in terms of diffusing diffusivity
models~\cite{chubynsky2014,chechkin2017,jain2017,jain2017b,tyagi2017,matse2017,jain2018,sposini2018,sposini2018b,grebenkov2019}.
It is only at time larger than $\tau$ that ordinary (Gaussian) Brownian motion is
recovered, with a diffusion coefficient $D_{\mathrm{av}}$.
Before giving an estimate of $\tau$ for our model (see next Section),
we study the early time non-Gaussianity in the full phase diagram $[N_{\mathrm{t}},\lambda]$, together
with its dependence on $\alpha$.

\begin{figure}[h!]
\begin{center} 
\includegraphics[width=.49\textwidth]{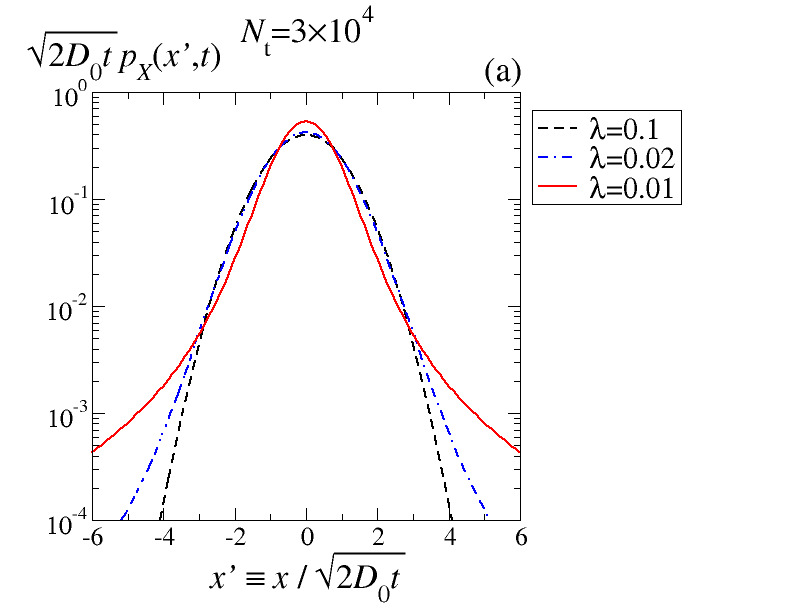}
\includegraphics[width=.49\textwidth]{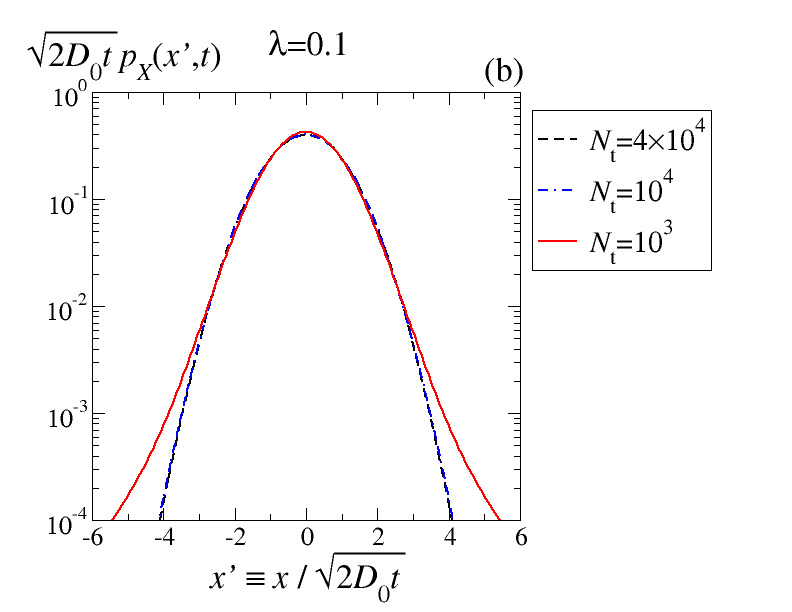}
\end{center}
\caption{PDF of the $x$-coordinate of $\boldsymbol{R}_G$ for
  $0\leq t\ll\tau$, at fixed $N_{\mathrm{t}}$ (a),
  and fixed $\lambda$ (b).
  The PDF is rescaled such that the variance is unity;
  recall that in
  a log-linear plot Gaussian PDFs have parabolic
  shape. In both cases, $\alpha=1$.
}
\label{fig:px}
\end{figure}

\begin{figure}[h!]
\begin{center} 
\includegraphics[width=.49\textwidth]{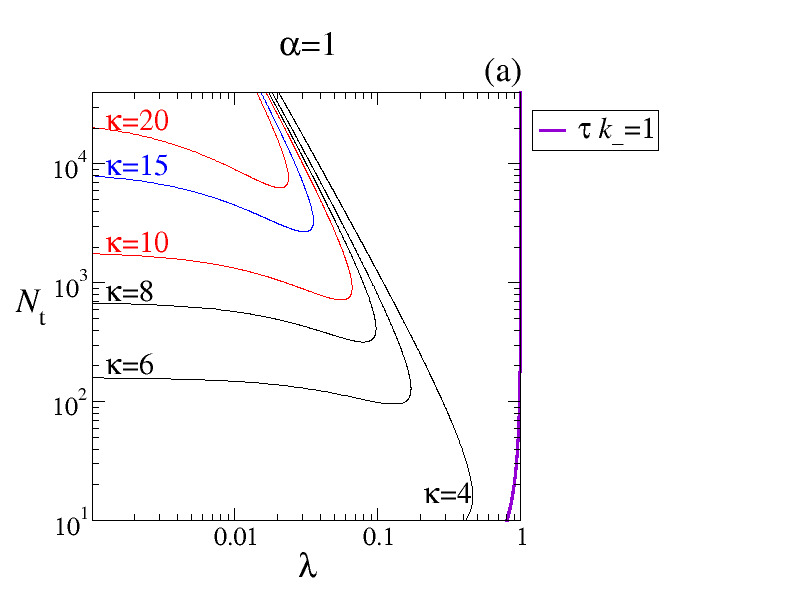}
\includegraphics[width=.49\textwidth]{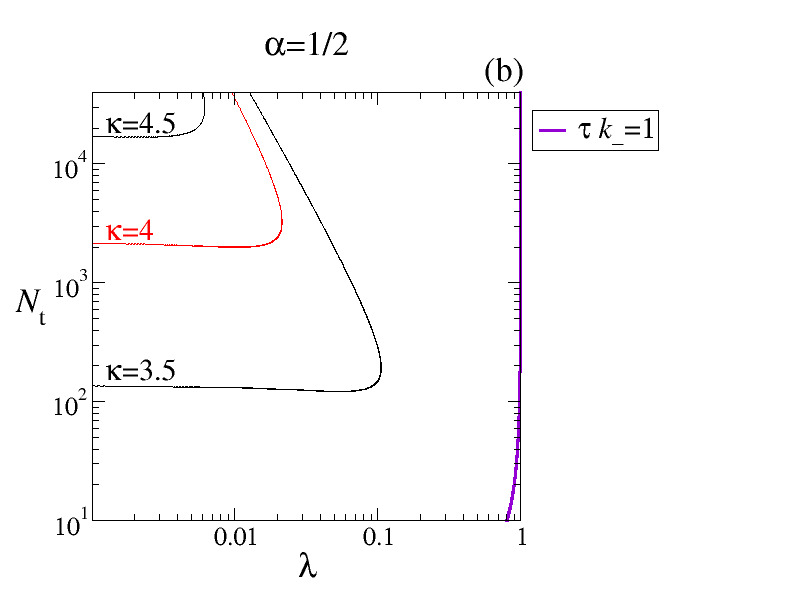}
\end{center}
\caption{Phase diagram of the early-stage non-Gaussianity. Labeled lines
  are the kurtosis  level curves. The thick, violet line at the right
  end of the plots corresponds to $\tau\,k_-$ (please refer to text for details).
}
\label{fig:phasediagram}
\end{figure}

The non-Gaussian behavior distinctive of 
$\boldsymbol{R}_G(t)$ at time $0\leq t\ll\tau$ can be properly characterized by referring to one of
its Cartesian coordinates, say $x$.
The PDF of the $x$-displacements takes the form 
\begin{equation}
  p_X(x,t)=\sum_{n=1}^{N_{\mathrm{t}}}P_{N}\left(\dfrac{D_0^\alpha}{D_n^\alpha}\right)
  \frac{
    \exp\left({-\frac{x^2}{4\pi\,D_nt}}\right)
  }{
    \sqrt{4\pi D_n t }
  }\,.
\label{ss}
\end{equation}
In Fig.~\ref{fig:px} we plot Eq.~\eqref{ss} for
$\alpha=1$ and different values of $\lambda$
and $N_{\mathrm{t}}$.
At first sight, non-Gaussianity increases with decreasing $N_{\mathrm{t}}$ and
and $\lambda$; below we however show that the behavior is
not monotonic.
To measure deviations from Gaussianity we consider the kurtosis of $p_X(x,t)$,
\begin{equation}
  \kappa\equiv
  \frac{
    \mathbb{E}\left[\left(X-\mathbb{E}[X]\right)^4\right]
  }{
    \left(\mathbb{E}\left[\left(X-\mathbb{E}[X]\right)^2\right] \right)^2
  }\,
\end{equation}
($\kappa=3$ for any Gaussian variable).
In our case it is straightforward to see that
\begin{equation}
  \kappa
  =3\,\dfrac{\mathbb{E}\left[D^2\right]}{\left(\mathbb{E}\left[D\right]\right)^2}
  =3\,\dfrac{\mathbb{E}\left[N^{-2\alpha}\right]}{\left(\mathbb{E}\left[N^{-\alpha}\right]\right)^2}\,,
\end{equation}
independently of $D_0$.
Notice instead the strong dependence of $\kappa$ from $\alpha$;
moreover,
$\kappa>3$ (positive excess kurtosis or leptokurtic PDF).
In order to illustrate regions
of more pronounced non-Gaussianity  and to discuss their dependence on
$\alpha$ in 
Fig.~\ref{fig:phasediagram} we draw the kurtosis level curves within a
$(\lambda,N_{\mathrm{t}})$-phase diagram
%lines separating regions of
%non-Gaussian (left) and Gaussian (right) behavior.
%In some cases (e.g., $100\%$ excess kurtosis), a
%minimum value of $N_{\mathrm{t}}$ is necessary to overcome the
%threshold.
Note that, for a given pair $(N_{\mathrm{t}},\lambda)$,
higher values of the exponent $\alpha$ give rise to  larger kurtosis
(compare Figs.~\ref{fig:phasediagram} a and b).

\begin{figure}[h!]
\begin{center} 
\includegraphics[width=.49\textwidth]{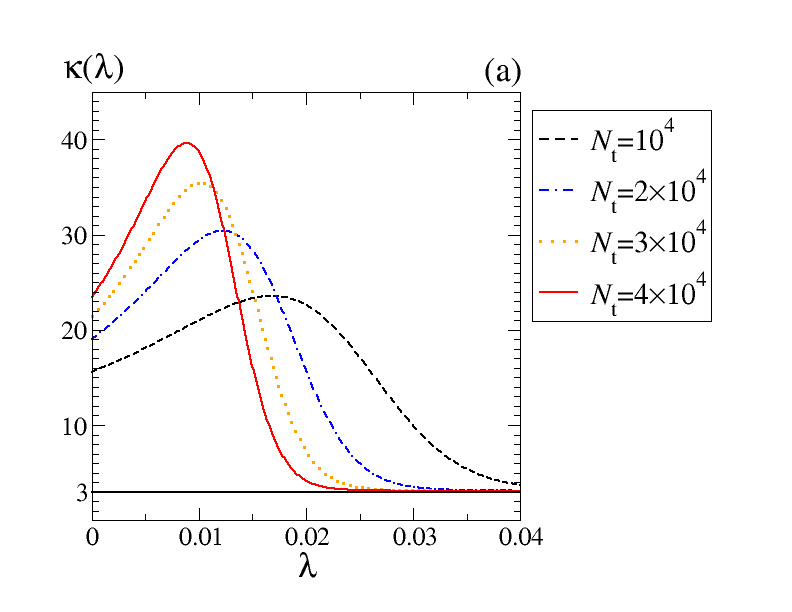}
\includegraphics[width=.49\textwidth]{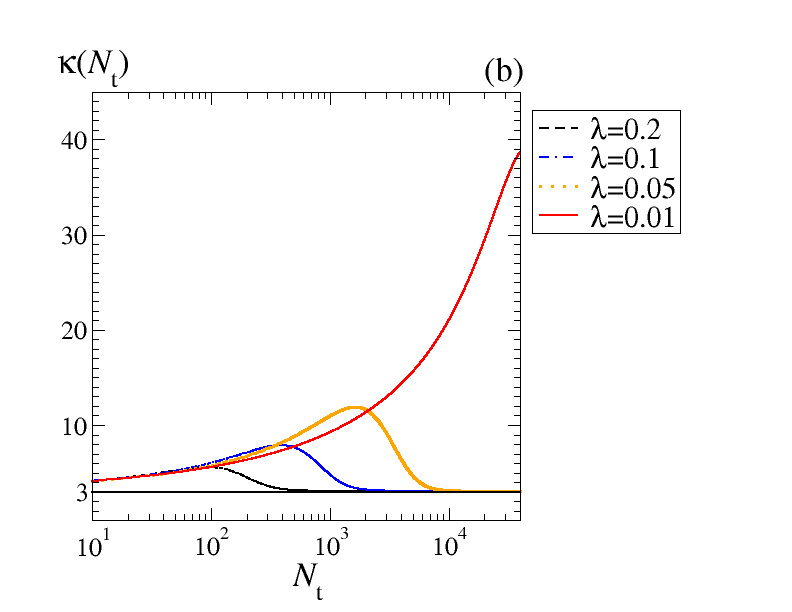}
\end{center}
\caption{Kurtosis as a function of: (a) $\lambda$;  (b) 
  $N_{\mathrm{t}}$. In both cases, $\alpha=1$.
}
\label{fig:maximumk}
\end{figure}

\begin{figure}[h!]
\begin{center} 
\includegraphics[width=.49\textwidth]{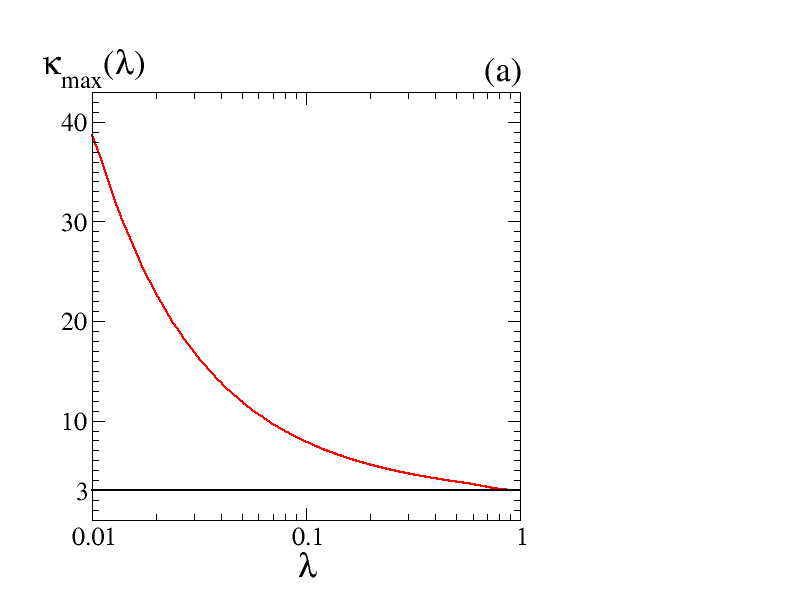}
\includegraphics[width=.49\textwidth]{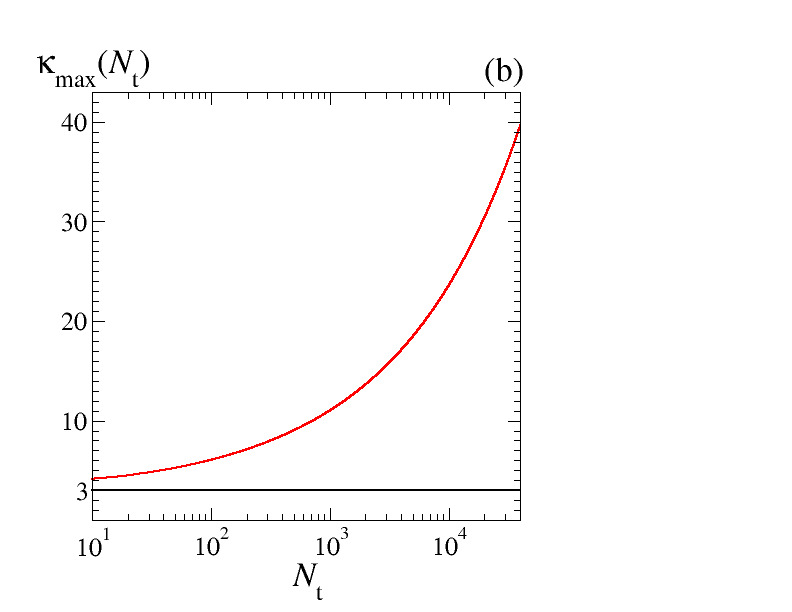}
\end{center}
\caption{Maximum kurtosis as a function of: (a) $\lambda$;  (b) 
  $N_{\mathrm{t}}$. In both cases, $\alpha=1$.
}
\label{fig:kmax}
\end{figure}

As quoted, by looking
at the plots in Fig.~\ref{fig:px} one may expect the
kurtosis to steadily increase by decreasing
$\lambda$ and $N_{\mathrm{t}}$.
The structure of the phase diagram implies instead the existence of
a maximum kurtosis, both at given $\lambda$ and $N_{\mathrm{t}}$.
This is highlighted in Fig.~\ref{fig:maximumk}.
Albeit within a small portion of the phase space, the maximum
kurtosis can be extremely high, as reported in Fig.~\ref{fig:kmax};
for instance, $k_{\mathrm{max}}\simeq40$ corresponds to an average polymer size
of order $N_{\mathrm{eq}}\simeq350$ 
with $N_{\mathrm{t}}\simeq10^4$.

\section{Crossover to Brownian, Gaussian diffusion}
\label{section_crossover}
The stationary distribution in Eq.~\eqref{eq_p_N_master} is exact, but it does
not provide information about the decay time-scale $\tau$ of initial
conditions for the process $N(t)$.
To get such an insight, we next workout a continuous
approximation for the polymerization process. 
In the gedankenexperiment reported above, $\tau$ is the persistence time
scale of the randomly chosen initial diffusion coefficient for
$\boldsymbol{R}_G$, corresponding in turn to the typical duration of
the leptokurtic PDF for the diffusion of the center of mass. 

\begin{figure}[h!]
\begin{center}
\includegraphics[width=.49\textwidth]{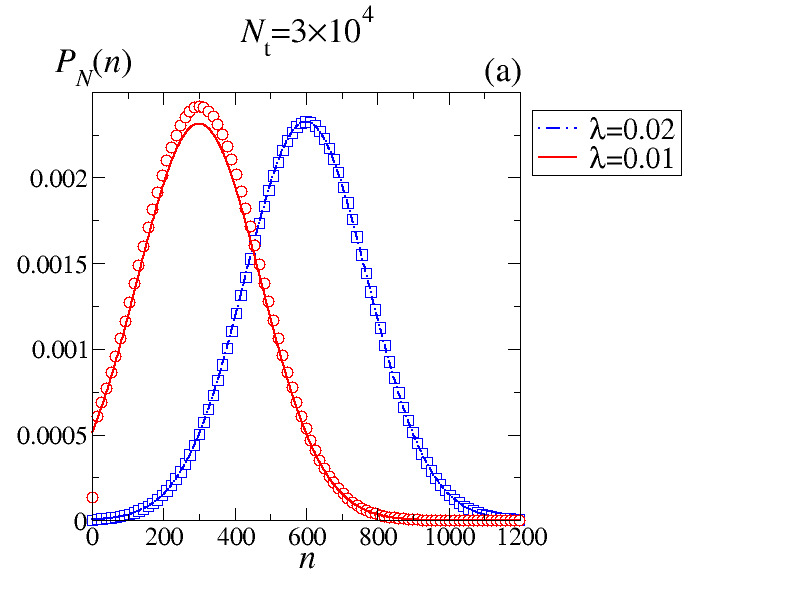}
\includegraphics[width=.49\textwidth]{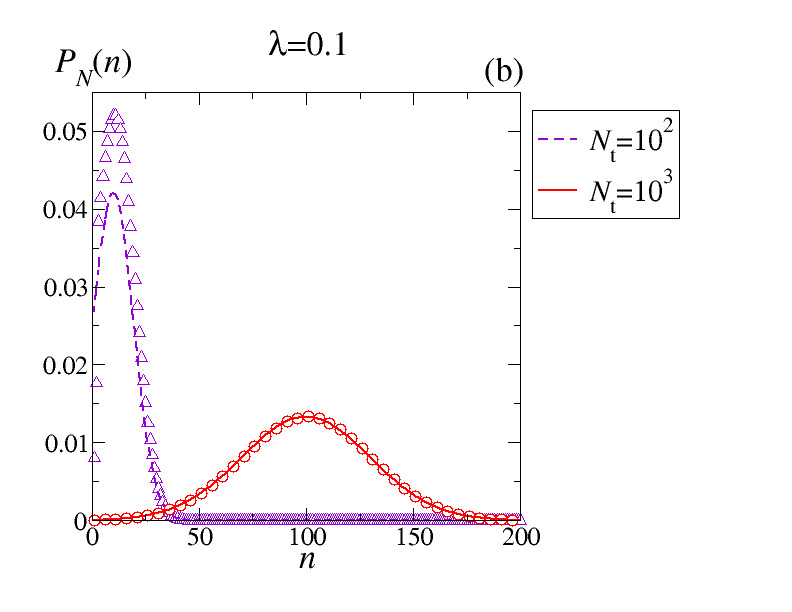}
\end{center}
\caption{Stationary PDF of the polymerization process.
  Comparison between the exact PDF in
  Eq.~\eqref{eq_p_N_master} (symbols) and the continuous, weak noise
  approximation associated to Eq.~\eqref{eq_N_stationary} (curves).
  Values for the parameters $N_{\mathrm{t}}$ and $\lambda$ have been
  chosen to facilitate comparison with
  Fig.~\ref{fig:px}. Specifically, continuous red curves
  correspond to choices in Fig.~\ref{fig:px}.
  Decreasing $\lambda$ at fixed $N_{\mathrm{t}}$ (a), or
  decreasing $N_{\mathrm{t}}$ at fixed $\lambda$ (b) the weak noise
  approximation breaks down.
}\label{fig:pN}
\end{figure}

We start by noticing that around equilibrium,  for $N_{\mathrm{t}}\gg1$ and
$N_{\mathrm{eq}}\gg M_{\mathrm{eq}}$ (large $\lambda)$,
$N(t)$ can  
be approximated as a continuous Markov process with Langevin
equation~\cite{gillespie1992} 
\begin{equation}
  \mathrm{d}N(t)
  =2\frac{k_+}{V}\left[N_{\mathrm{eq}}-N(t)\right]\,\mathrm{d}t
  +\sqrt{2\frac{k_+}{V}\left[2N_{\mathrm{t}}-N_{\mathrm{eq}}-N(t)\right]}\,\mathrm{d}B(t)\,,
  \label{eq_N}
\end{equation}
where $B(t)$ is a Wiener process (Brownian motion). 
Taking further advantage of the 
large $N_{\mathrm{eq}}$ assumption, we then introduce the rescaled
quantity $\widetilde{N}\equiv N/N_{\mathrm{eq}}$, obeying
\begin{equation}
  \mathrm{d}\widetilde{N}(t)
  =2\frac{k_+}{V}\left[1-\widetilde{N}(t)\right]\,\mathrm{d}t
  +\left(\frac{1}{N_{\mathrm{eq}}}\right)^{1/2}\sqrt{2\frac{k_+}{V}\left[2\frac{N_{\mathrm{t}}}{N_{\mathrm{eq}}}-1-\widetilde{N}(t)\right]}\,\mathrm{d}B(t)\,,
  \label{eq_N_rescaled}
\end{equation}
to which we may
apply the \textit{weak noise approximation}.
Indeed, one may straightforwardly prove~\cite{gillespie1992} that for
large $N_{\mathrm{eq}}$ Eq.~\eqref{eq_N_rescaled} is satisfied by  the
approximate solution
\begin{equation}
\widetilde{N}(t)\simeq
\widetilde{n}(t)+\left(\frac{1}{N_{\mathrm{eq}}}\right)^{1/2} Y(t)\,,
\end{equation}
with $\widetilde{n}(t)$ a deterministic process satisfying
\begin{equation}
  \frac{\mathrm{d}\widetilde{n}(t)}{\mathrm{d}t}
  =2\frac{k_+}{V}\left[1-\widetilde{n}(t)\right]\,,
\end{equation}
and $Y(t)$ the stochastic process defined by the Langevin equation
\begin{equation}
  \mathrm{d}Y(t)
  =-2\frac{k_+}{V}\,Y(t)\,\mathrm{d}t
  +\sqrt{2\frac{k_+}{V}\left[2\frac{N_{\mathrm{t}}}{N_{\mathrm{eq}}}-1-\widetilde{n}(t)\right]}\,\mathrm{d}B(t)\,.
\end{equation}
The solution of the deterministic process,
\begin{equation}
\widetilde{n}(t)=1+[\widetilde{n}(0)-1]\,\mathrm{e}^{-\frac{t}{\tau}}\,,
\end{equation}
asymptotically tends to $1$ with a characteristic decay time
\begin{equation}
  \tau\equiv\frac{V}{2k_+}
  =\frac{(1-\lambda)\,N_{\mathrm{t}}}{2k_-}\,\,.
  \label{eq_tau}
\end{equation}
Correspondingly, the long-time behavior of $Y(t)$ is that of an
Ornstein-Uhlenbeck process:
\begin{equation}
Y(t\to\infty)=\mathbb{N}\left(0,\frac{N_{\mathrm{t}}}{N_{\mathrm{eq}}}-1\right)\,,
\end{equation}
where $\mathbb{N}(\mu,\sigma^2)$ is a Gaussian variable with mean $\mu$
and variance $\sigma^2$. Hence, the stationary solution of
$\widetilde{N}$ is
\begin{equation}
\widetilde{N}(t\to\infty)=\mathbb{N}\left(1,\frac{M_{\mathrm{eq}}}{N_{\mathrm{eq}}^2}\right)\,.
\end{equation}
For the polymer size $N=\widetilde{N}\,N_{\mathrm{eq}}$, this implies
\begin{equation}
  N(t\to\infty)=\mathbb{N}\left(N_{\mathrm{eq}},M_{\mathrm{eq}}\right)\,.
  \label{eq_N_stationary}
\end{equation}
We thus appreciate that, to be self consistent, the continuous
approximation requires large values of $N_{\mathrm{t}}$ to blur out discreteness,
and $N_{\mathrm{eq}}\gg M_{\mathrm{eq}}$ so that the negative support of
the Gaussian PDF corresponds to a negligible probability.
Fig.~\ref{fig:pN} shows that when $N_{\mathrm{t}}$ and $\lambda$ are both
large the weak noise approximation of the  stationary
distribution $P_N(n)$ is almost indistinguishable from the exact solution. On
the other hand, decreasing either $N_{\mathrm{t}}$ or $\lambda$ 
the approximation fails concomitantly with the fact that the Gaussian
probability of negative $n$-values becomes significant.
Depending on the specific cut in phase-space, the approximation may
or may not work well in correspondence to the maximum kurtosis (compare
red full lines in Figs.~\ref{fig:pN} a and b).  

When applicable, the important result conveyed by the continuous, weak
noise approximation is that through Eq.~\eqref{eq_tau} it establishes
the time scale of the decay of the autocorrelation of $N(t)$.
To Fig.~\ref{fig:phasediagram}, we thus added the line
\begin{equation}
\tau\,k_-=1\,.
\end{equation}
As it depends on the dissociation rate constant specific of the chosen
polymer, this line has to be understood qualitatively: according to
our estimation, the farther left of
the line, the longer lasts the Brownian yet non-Gaussian diffusion stage.

\section{Conclusions}
We have been able to analytically characterize the stochastic motion of the center
of mass of a fluctuating filament undergoing a simple polymerization
process.
Depending on experimentally accessible parameters such as the
the total monomers in the solution $N_{\mathrm{t}}$ and the system volume
$V$ (equivalently, the fraction $\lambda$ of total  monomers composing the
filament in equilibrium), the center of mass displays at early times a
Brownian, yet non-Gaussian,
diffusion. To our knowledge, this is one of the first example in which
this anomalous behavior is directly linked  to a microscopic prototype: 
the effect originates from the fluctuations of $N$ (due to
polymerization)
and from the relation $D(N) =D_0/N^{\alpha}$ which 
distinguishes many microscopic models of polymeric diffusion.
By studying the kurtosis of the 
early-time displacement PDF along the $x$-coordinate we quantified deviations
from Gaussian behavior in the phase diagram
$(\lambda,N_{\mathrm{t}})$, highlighting  the dependence on the exponent
$\alpha$. Remarkably, the kurtosis is not monotonic and displays a maximum at either
$\lambda$  or $N_{\mathrm{t}}$ fixed.
Finally, on the basis of a continuum (weak noise) approximation for the
stochastic process $N(t)$, we put forward an estimation for
the time $\tau(\lambda,N_{\mathrm{t}})$ at which the anomalous
behavior crosses over to ordinary Brownian motion.
Since the weak noise approximation is not applicable in the whole
$(\lambda,N_{\mathrm{t}})$ phase diagram, and also in view of the
non-monotonic behavior of the kurtosis, further studies approaching
the determination of $\tau$ are welcome. 

In parallel with the analytical results, we proposed a
{\it gedankenexperiment} in which the anomalous behavior could be
detected. As a further perspective, we may notice that if we
shift the focus on the diffusion of a
tagged monomer (in place of the center of
mass of the polymer), in the early stage of the process a
{\it subdiffusive} behavior coupled to non-Gaussianity is expected to be
observed, with a crossover to a Brownian regime at the
Rouse time~\cite{doi1988}. This analysis is intended to be the subject
of future work.  

In conclusion, we believe that this work provides a valuable analytical
backdrop to Brownian yet non-Gaussian diffusion,
a fascinating phenomenon reported to occur in many physical systems.
To fully understand this anomalous behavior, it is essential to ground
it on a microscopic spring.
This is the case for the presented model, but we are
confident than others more will come along these lines.

\section*{Appendix}
The stationary distribution
$P_N(n)\equiv\lim_{t\to\infty}P_N(n,t|n_0,t_0)$ can be
obtained by putting $\partial_t P_N(n,t|n_0,t_0)= 0$ in Eq.~\eqref{eq_master}, 
\begin{equation}
    W_-(n+1)\,P_N(n+1) =  [W_+(n)+W_-(n)]\,P_N(n)-W_+(n-1)\, P_N(n-1)\,,
  \label{eq_master_stat}
\end{equation}
and then solving recursively.
Let us first consider the case $N=1$.
Since with $N_{\mathrm{t}}>0$ we always have at least a polymer of
size $1$, $P_N(0)=0$. Moreover, as there are no bonds to be broken down
with a polymer of size one, $W_-(1)=0$.
This gives
\begin{equation}
    W_-(2)\,P_N(2) =  W_+(1)\,P_N(1).
  \label{eq_rec_1}
\end{equation}
With $n=2$, Eq.~(\ref{eq_master_stat})  becomes
\begin{equation}
    W_-(3)\,P_N(3) =  [W_+(2)+W_-(2)]\,P_N(2)-W_+(1)\,P_N(1)
  \label{eq_rec_2}
\end{equation}
and plugging Eq.~(\ref{eq_rec_1}) into Eq.~(\ref{eq_rec_2}) we get
\begin{equation}
    W_-(3)\,P_N(3) =  \frac{W_+(2)}{W_-(2)}W_+(1)\,P_N(1). 
  \label{eq_rec_3}
\end{equation}
Since  
\begin{equation}
    W_-(4)\,P_N(4) =  \frac{W_+(3)}{W_-(3)}\frac{W_+(2)}{W_-(2)}W_+(1)\,P_N(1),. 
  \label{eq_rec_4}
\end{equation}
one can  assume for any $n>2$
\begin{equation}
    W_-(n)\,P_N(n) =  \left (\prod_{n'=2}^{n-1}\frac{W_+(n')}{W_-(n'))} \right )W_+(1)\,P_N(1)\,, 
  \label{eq_rec_M}
\end{equation}
and prove that the same holds with $n+1$. Indeed,
\begin{eqnarray}
  W_-(n+1)\,P_N(n+1) &=&   [W_+(n)+W_-(n)]\,P_N(n)-W_+(n-1)\,P_N(n-1)\nonumber \\ 
  &=&\left[
    \left( \frac{W_+(n)}{W_-(n)} +1 \right)
    \left(\prod_{n'=2}^{n-1}\frac{W_+(n')}{W_-(n')} \right)
    -\frac{W_+(n-1)}{W_-(n-1)}
    \left(\prod_{n'=2}^{n-2}\frac{W_+(n')}{W_-(n')} \right)
    \right] W_+(1)\,P_N(1)
  \nonumber\\
  &=& \left (\prod_{n'=2}^{n}\frac{W_+(n')}{W_-(n')} \right ) W_+(1)\,P_N(1)\,.
  \label{eq_rec_M+1}
\end{eqnarray}
As the normalization condition $\sum_{n=1}^{N_{\mathrm{t}}} P_N(n) = 1$  gives
\begin{equation}
  P_N(1)
  +\frac{W_+(1)}{W_-(2)}\,P_N(1)
  +\sum_{n=3}^{N_{\mathrm{t}}} \frac{1}{W_-(n)}
  \left (\prod_{n'=2}^{n-1}\frac{W_+(n')}{W_-(n'))} \right
  )W_+(1)\,P_N(1)
  =1\,,
\end{equation}
or
\begin{equation}
  W_+(1)\,P_N(1)
  = \frac{1}
  {
    \frac{1}{W_+(1)}
    +\frac{1}{W_-(2)}
    +\sum_{n=3}^{N_{\mathrm{t}}} \frac{1}{W_-(n)} \left
    (\prod_{n'=2}^{n-1}\frac{W_+(n')}{W_-(n')} \right )
  }\,,
\end{equation}
we now get 
\begin{equation}
  \begin{array}{lll}
    P_N(1)
    &=&
    \displaystyle
    \frac{
      \frac{1}{W_+(1)}
    }{
      \frac{1}{W_+(1)}
      +\frac{1}{W_-(2)}
      +\sum_{n=3}^{N_{\mathrm{t}}} \frac{1}{W_-(n)} \left
      (\prod_{n'=2}^{n-1}\frac{W_+(n')}{W_-(n')} \right )
    }\\
    &&\\
    P_N(2)
    &=&
    \displaystyle
    \frac{
      \frac{1}{W_-(2)}
    }{
      \frac{1}{W_+(1)}
      +\frac{1}{W_-(2)}
      +\sum_{n=3}^{N_{\mathrm{t}}} \frac{1}{W_-(n)} \left
      (\prod_{n'=2}^{n-1}\frac{W_+(n')}{W_-(n')} \right )
    }\\
    &&\\
    P_N(n)
    &=&
    \displaystyle
    \frac{
      \frac{1}{W_-(n)}
      \;\left (\prod_{n'=2}^{n-1}\frac{W_+(n')}{W_-(n')}
      \right)
    }{
      \frac{1}{W_+(1)}
      +\frac{1}{W_-(2)}
      +\sum_{n'=3}^{N_{\mathrm{t}}} \frac{1}{W_-(n')} \left
      (\prod_{n''=2}^{n'-1}\frac{W_+(n'')}{W_-(n'')} \right )
    }\quad(3\leq n\leq N_{\mathrm{t}})
    \end{array}\,.  
\label{staz_distr_gen}
\end{equation}
The result in Eq.~(\ref{staz_distr_gen}) is rather general, as the transition rates 
are not specified.
Applying the stepping functions in Eq.~\eqref{eq_stepping}, we have
\begin{equation}
  \begin{array}{rll}
    \dfrac{1}{W_+(1)}
    &=&
    \dfrac{V}{2k_+\,(N_{\mathrm{t}}-1)}
    =\dfrac{(1-\lambda)\,N_{\mathrm{t}}}{2k_-\,(N_{\mathrm{t}}-1)}
    \\
    &&\\
    \dfrac{1}{W_-(2)}
    &=&
    \dfrac{1}{k_-}
    \\
    &&\\
    \displaystyle
    \frac{1}{W_-(n)}
    \left(\prod_{n'=2}^{n-1}\frac{W_+(n')}{W_-(n')} \right)    
    &=&
    \displaystyle
    \frac{1}{k_-}
    \;\frac{2k_+\,(N_{\mathrm{t}}-2)}{k_-\,V}
    \;\prod_{n'=3}^{n-1}\frac{k_+\,(N_{\mathrm{t}}-n')}{k_-\,V}
    =\frac{2}{k_-}
    \;\left(\frac{k_+}{k_-\,V}\right)^{n-2}
    \;\frac{(N_{\mathrm{t}}-2)!}{(N_{\mathrm{t}}-n)!}
    \\
    &&\\
    &=&
    \displaystyle
    \frac{2}{k_-}
    \;\left[(1-\lambda)\,N_{\mathrm{t}}\right]^{2-n}
    \;\frac{(N_{\mathrm{t}}-2)!}{(N_{\mathrm{t}}-n)!}
    \\
    &&\\
    &=&
    \displaystyle
    \frac{2}{k_-}
    \;\frac{(N_{\mathrm{t}}-2)!}{\left[(1-\lambda)\,N_{\mathrm{t}}\right]^{N_{\mathrm{t}}-2}}
    \;\frac{\left[(1-\lambda)\,N_{\mathrm{t}}\right]^{N_{\mathrm{t}}-n}}{(N_{\mathrm{t}}-n)!}
    \quad(3\leq n\leq N_{\mathrm{t}})
  \end{array}\,.
  \nonumber
\end{equation}
Taking advantage of the identity
\begin{equation}
  \sum_{n=0}^{N_{{\mathrm{t}}}}\frac{\left[(1-\lambda)\,N_{\mathrm{t}}\right]^{N_{\mathrm{t}}-n}}{(N_{\mathrm{t}}-n)!}
  =\frac{\Gamma(N_{\mathrm{t}}+1,(1-\lambda)\,N_{\mathrm{t}})}{\Gamma(N_{\mathrm{t}}+1)}
  \;\mathrm{e}^{(1-\lambda)\,N_{\mathrm{t}}}\,,
\end{equation}
$\Gamma(\cdot,\cdot)$ being the upper incomplete gamma
function~\cite{abramowitz1965},
and $\Gamma(\cdot)$ the Euler gamma
function, an explicit expression for the normalization factor is
\begin{eqnarray}
  \mathcal{N}(N_{\mathrm{t}},\lambda)
  &\equiv&
  \frac{1}{W_+(1)}
  +\frac{1}{W_-(2)}
  +\sum_{n=3}^{N_{\mathrm{t}}} \frac{1}{W_-(n)} \left
  (\prod_{n'=2}^{n-1}\frac{W_+(n')}{W_-(n')} \right )
  \\
  &=&
  \dfrac{(1-\lambda)\,N_{\mathrm{t}}}{2\,(N_{\mathrm{t}}-1)}
  +1
  -2\frac{(1-\lambda)^2\,N_{\mathrm{t}}}{N_{\mathrm{t}}-1}
  -2\frac{(1-\lambda)\,N_{\mathrm{t}}}{N_{\mathrm{t}}-1}
  -2
  \nonumber\\
  &&
  +\frac{2\,(N_{\mathrm{t}}-2)!}{\left[(1-\lambda)\,N_{\mathrm{t}}\right]^{N_{\mathrm{t}}-2}}
  \;\frac{\Gamma(N_{\mathrm{t}}+1,(1-\lambda)\,N_{\mathrm{t}})}{\Gamma(N_{\mathrm{t}}+1)}
  \;\mathrm{e}^{(1-\lambda)\,N_{\mathrm{t}}}
  \nonumber\\
  &=&
  \frac{N_{\mathrm{t}}\,[(11-4\lambda)\,\lambda-9]+2}{2(N_{\mathrm{t}}-1)}
  \nonumber\\
  &&
  +\frac{2\,(N_{\mathrm{t}}-2)!}{\left[(1-\lambda)\,N_{\mathrm{t}}\right]^{N_{\mathrm{t}}-2}}
  \;\frac{\Gamma(N_{\mathrm{t}}+1,(1-\lambda)\,N_{\mathrm{t}})}{\Gamma(N_{\mathrm{t}}+1)}
  \;\mathrm{e}^{(1-\lambda)\,N_{\mathrm{t}}}\,.
  \end{eqnarray}
Putting things together, the exact stationary solution of Eq.~\eqref{eq_master} is
given by Eq.~\eqref{eq_p_N_master}.

\newpage
\section*{Figure captions}

% For Original Research articles, please note that the Material and
% Methods section can be placed in any of the following ways: before
% Results, before Discussion or after Discussion.

\clearpage
\section*{Conflict of Interest Statement}
%All financial, commercial or other relationships that might be perceived by the academic community as representing a potential conflict of interest must be disclosed. If no such relationship exists, authors will be asked to confirm the following statement: 

The authors declare that the research was conducted in the absence of any commercial or financial relationships that could be construed as a potential conflict of interest.

\section*{Author Contributions}
All authors equally contributed to the present article.

\section*{Acknowledgments}
The authors would like to thank M. Baiesi, G. Falasco,
and A.L. Stella for useful discussions. 
FB and FS acknowledge financial support from a 2019 PRD project of the
Physics and Astronomy Department of the University of Padova, Italy.

\bibliographystyle{apsrev4-1}
\bibliography{draft_bibliography}

%%% Make sure to upload the bib file along with the tex file and PDF
%%% Please see the test.bib file for some examples of references

\end{document}